# First-Principle Study of Dislocation Slips in Impurity Doped Graphene

Fanchao Meng, Bin Ouyang, and Jun Song[*]

Department of Mining and Materials Engineering,

McGill University, Montréal, Québec H3A 0C5, Canada

[*] Author to whom correspondence should be addressed. Email: jun.song2@mcgill.ca Tel.: +1 (514) 398-4592 Fax: +1 (514) 398-4492

**ABSTRACT**

Employing density-functional theory (DFT) calculations, the generalized-stacking-fault energy (GSFE) curves along two crystallographic slips, glide and shuffle, for both pristine graphene and impurity of boron (B) or nitrogen (N) doped graphene were examined. The effects of B and N doping on the GSFE were clarified and correlated with local electron interactions and bonding configurations. The GSFE data were then used to analyze dislocation dipole and core structure, and subsequently combined with the Peierls-Nabarro (P-N) model to examine the role of doping on several key characteristics of dislocations in graphene. We showed that the GSFE curve may be significantly altered by the presence of dopants, which subsequently leads to profound modulations of dislocation properties, such as increasing spontaneous pair-annihilation distance and reducing resistance to dislocation slip. Our results indicate that doping can play an important role in controlling dislocation density and microscopic plasticity in graphene, thereby providing critical insights for dopant-mediated defect engineering in graphene.

## I. INTRODUCTION

Graphene, a strictly two-dimensional monolayer material with atoms arranged in a hexagonal honeycomb lattice, exhibits many extraordinary physical and electronic properties[1-6] and attracts enormous research efforts.[7-13] In particular, owing to its exceptional mechanical properties, graphene promises numerous possibilities in applications including composites,[14-16] pressure barriers,[17] filters,[18] and sensors,[19-22] among others. However, various lattice defects, e.g., vacancies,[23-26] Stone-Wales (SW) defects,[24-27] dislocations[23, 28-31] and grain boundaries (GBs),[32-34] will emerge accompanying the fabrication or growth of graphene. Those structural singularities are shown to degrade the mechanical properties of graphene,[35-43] posting a significant limitation on the applications of graphene.

One important class of lattice defects in graphene is dislocations. In a monolayer graphene, the dislocation is edge in nature. It has the form of pentagon-heptagon pair[44] and is thermodynamically stable.[45-46] Dislocations are also main constituents for many GBs[30-31] and thus play a key role in determining the strength and fracture behaviors of polycrystalline graphene.[37-42] As a result the knowledge of dislocations is of fundamental importance to the understanding of deformation and failure mechanisms in graphene. Besides lattice defects like dislocations, another category of defects often present in graphene is impurities. Impurities can come from various chemical processes during the synthesis of graphene[33, 47] or intentionally introduced as dopants to modify the electronic properties of graphene.[48-51] Among various impurity atoms, B and N atoms are of particular interest given their similarities to C atom[52-54] and their importance as dopants (i.e., B and N are common *p*-type and *n*-type dopants respectively[55-58]) to tune the electronic properties of graphene. With dislocations and impurities coexisting in graphene, they may interact to further influence the properties of graphene in addition to their own individual effects on graphene. In particular, the impurity atoms, often of different atomic radii from the C atom, will tend to migrate towards dislocations to reduce the overall strain energy. The segregation will lead to clouds of impurities at dislocations, thus necessarily modify the mechanics and dynamics of dislocations.

In this paper, we investigate the slip properties of dislocations in graphene and the effects of B and N impurities within the framework of the generalized-stacking-fault energy (GSFE) curve proposed by Vitek.[59-60] The GSFE curve yields critical information of the energy cost associated with the slip/shearing of lattice during dislocation motions. It also provides important inputs for the Peierls-Nabarro (P-N) model that enables continuum examination of the dislocation characteristics.[61-67] The GSFE curves along two different slip directions, with or without dopants (i.e., B and/or N) in graphene, were computed using first-principle calculations. The influence of impurities on the dislocation characteristics and slip mechanisms were then examined analytically within the P-N model. In the end the implications of B and N doping on plastic deformation in graphene were discussed.

## II. COMPUTATIONAL METHOD

To obtain the GSFE curves, spin polarized density-functional theory (DFT)[68-69] calculations were performed using the Vienna *Ab-initio* Simulation Package (VASP).[70] The simulation cells for the GSFE calculations are illustrated in Figure 1. In each simulation cell, a zigzag graphene nanoribbon of width around 29 Å and with edges passivated by hydrogen atoms, is enclosed. The lattice constant used to construct the nanoribbon is 2.46 Å, the one obtained from the perfect monolayer graphene, in agreement with values previously reported.[71-73] The simulation cell's dimension along the armchair direction is chosen as 49 Å and its dimension perpendicular to the monolayer is set as 15 Å in order to eliminate the interlayer interactions, whilst the cell's dimension along the zigzag direction varies (see below) depending on the simulation. Two types of crystallographic slips are considered, illustrated in Figure 1, termed as glide and shuffle slips according to Ref. 74, with the slip lines indicated by dashed black and dash-dot red lines respectively. To compute the GSFE curve, the atoms on one side of the slip line (cf. Figure 1) are displaced with respect to the other side, and the attendant energy cost per unit length, i.e., the GSFE $\gamma$ associated with the slip displacement $\delta$ is monitored. The slip process and the following relaxation are in accordance with previous studies.[59, 66, 74] To examine the effects of B and/or N impurity atoms on the GSFE, they are introduced to substitute C atoms either along the slip line or along the atomic row immediately neighboring the slip line. For simplicity we use acronyms *AS* and *NS* as superscripts to indicate dopants along and immediately neighboring the slip line respectively. We define the line concentration of the impurity $\alpha$ ($\alpha$ = B or N), denoted as $c_\alpha$, as the line density of the impurity normalized by the corresponding line density of C in the pristine graphene. Depending on the amount of impurities within the system, different dimension along the zigzag direction is used for the simulation cell, ranging from 4.9 Å to 24.6 Å.[75]

In the DFT calculation, the exchange correlation functions are approximated by generalized gradient approximation (GGA) of Perdew, Burke, and Ernzerhof (PBE).[76] The electron-ion interactions for elements C, B, and N are described using the projector augmented wave (PAW) method.[77] The plane wave basis cut off of 500 eV is used for all calculations. The ion positions were relaxed with the force tolerance being 0.03 eV/Å.[78]

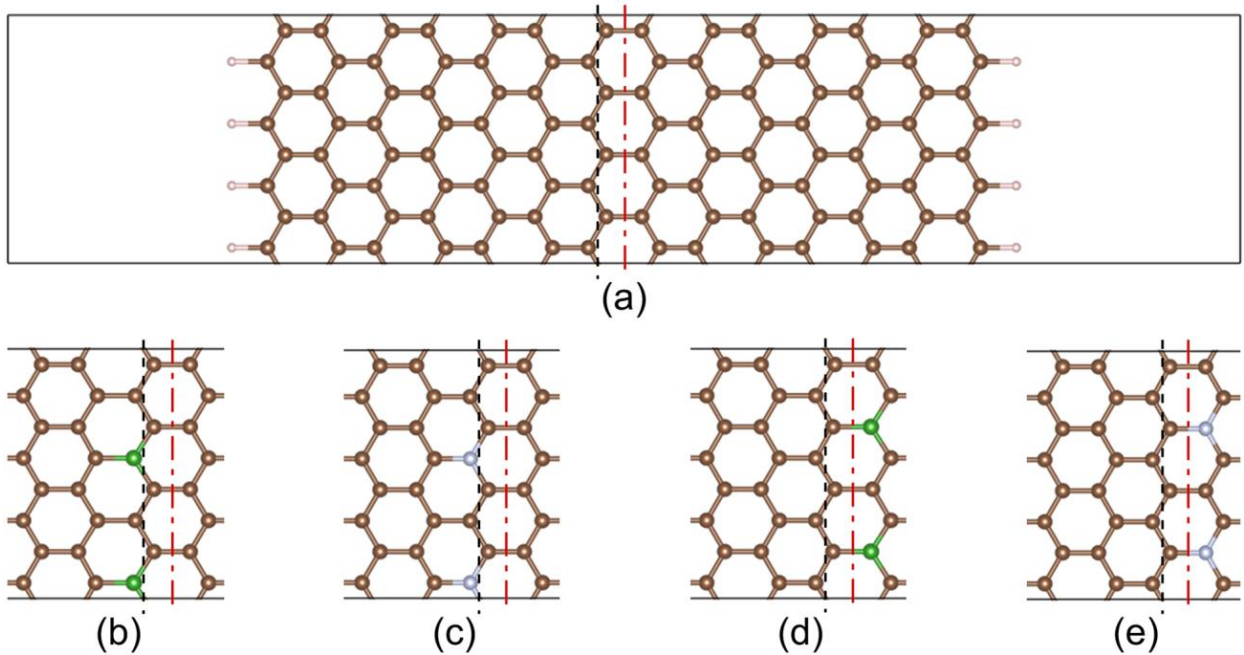

**Figure 1.** The simulation cells for (a) the pristine graphene, and four representative configurations of graphene doped by impurities, i.e., the configurations where the graphene is doped with (b) B of concentration $c_B = 0.5$ and (c) N of concentration of $c_N = 0.5$ along the glide slip line (black dash) or equivalently along the row of atoms immediately neighboring the shuffle slip line (red dash-dot), and (d) B of concentration $c_B = 0.5$ and (e) N of concentration $c_N = 0.5$ along the shuffle slip line (red dash-dot) or equivalently along the row of atoms immediately neighboring the glide slip line (black dash). The free-edges of the graphene sheet are passivated by hydrogen atoms. The C, B, N, and H atoms are colored by brown, green, silver and rose respectively.

## III. RESULTS AND DISCUSSIONS

### 3.1. GSFE curves along glide direction

The GSFE curve for the glide slip in pristine graphene is plotted in Figure 2a, showing a wide plateau between $\delta = 0.25b$ and $\delta = 0.75b$, with b denoting the magnitude of the Burgers vector. The curve also exhibits a local minimum, i.e., the meta-stable stacking fault energy $\gamma_{sf}$, at $\delta = 0.5b$, that is slightly lower than the unstable stacking fault energy $\gamma_{usf}$, suggesting the possibility of a full dislocation dissociating into two partials with a weak tendency. Figure 2a also shows two representative GSFE curves with $B^{AS}$ and $N^{AS}$ doping respectively, from which we note that both $B^{AS}$ and $N^{AS}$ doping lead to overall decrease of GSFE. In addition we see that doping noticeably modifies the shape of the GSFE curve, and it appears that $N^{AS}$ doping amplifies while $B^{AS}$ doping moderates or even eliminates the local GSFE valley at $\delta = 0.5b$. The effects of doping are further illustrated in Figures 2b and 2c where $\gamma_{usf}$, $\gamma_{sf}$, and $\gamma_{usf} - \gamma_{sf}$, are plotted as functions of the dopant concentration, $c_\alpha$, showing that both $\gamma_{usf}$ and $\gamma_{sf}$ monotonically decrease as the dopant concentration increases, with the reduction being more pronounced in the case of $B^{AS}$ doping. Meanwhile $\gamma_{usf} - \gamma_{sf}$ overall increases as $c_N$ increases but quickly approaches zero as $c_B$ increases. At high concentrations of $B^{AS}$ doping (i.e., $c_B > 20\%$, indicated by solid triangles in Figure 2c), $\gamma_{usf} - \gamma_{sf}$ becomes zero, indicating that the meta-stable stacking fault no longer exists.

The effects of dopants when they reside along the atomic row immediately neighboring the glide slip line are also shown in Figure 2. As illustrated in Figure 2d, $N^{NS}$ doping leads to a slight increase in the GSFE while $B^{NS}$ doping decreases the GFSE, yet in both cases the shape of the GSFE curve remains largely unaltered. These trends are also reflected in the $\gamma_{usf}$ and $\gamma_{sf}$ plots in Figure 2e, where $\gamma_{usf}$ and $\gamma_{sf}$ monotonic climb and decline as $c_N$ and $c_B$ increase respectively.

For the quantity, $\gamma_{usf}$ - $\gamma_{sf}$ (cf. Figure 2f), it exhibits a slight increase with N$^{NS}$ doping but quickly diminishes with B$^{NS}$ doping, being somewhat similar as the ones observed in the case with dopants along the slip line.

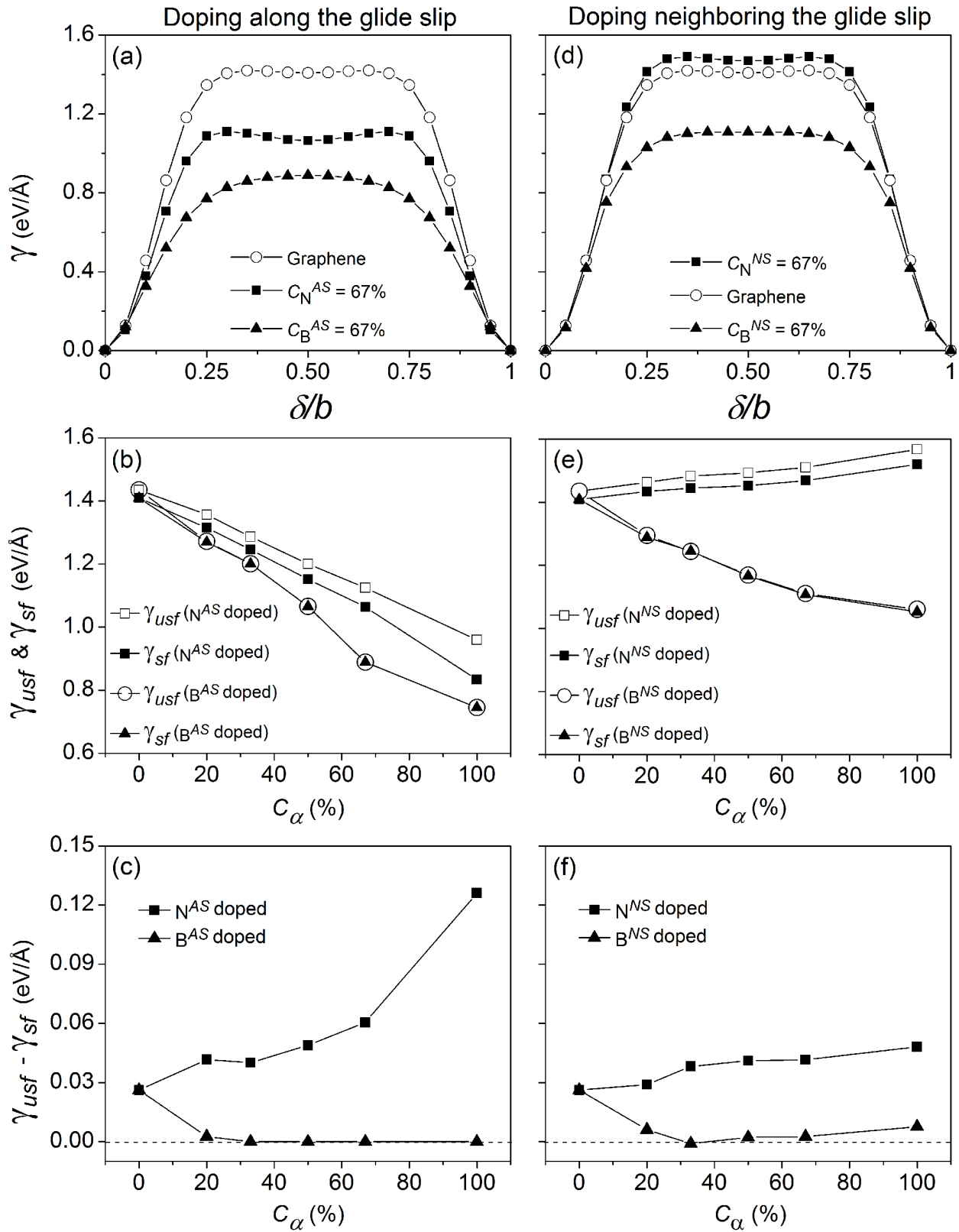

**Figure 2.** Sample GSFE curves for graphene (open circles) and graphene with B (solid triangles) and N (solid squares) doping along (a) the glide slip line and (d) atomic row immediately neighboring the glide

slip line. The unstable and stable stacking fault energies, $\gamma_{usf}$ (open symbols) and $\gamma_{sf}$ (solid symbols), and their difference, $\gamma_{usf}$ - $\gamma_{sf}$, as functions of the dopant concentration are shown in (b) and (c) respectively for doping along the glide slip line, and (e) and (f) respectively for doping along atomic row immediately neighboring the glide slip line. The acronyms *AS* and *NS* indicate dopants along and immediately neighboring the slip line respectively.

To understand the effects of B and N doping on the GSFE, the formation energies (denoted as $\Omega$) to incorporate them into a pristine graphene sheet were calculated:[79-80]

$$\Omega = E[X] - E[G] - \mu_X + \mu_C, \qquad (1)$$

where $E[X]$ is the total energy of the supercell with one impurity atom (X = B or N), and $E[G]$ is the total energy of the graphene supercell. $\mu_X$ and $\mu_C$ are the chemical potentials for the impurity (X = B or N) and C respectively, obtained from α-boron bulk, $N_2$ molecule and pristine graphene sheet.[79] The formation energies for B and N are obtained to be 1.00 eV and 0.63 eV respectively, both being positive and in agreement with the values previously reported.[79, 81-82] This suggests the order of C-C > C-N > C-B in bond strength. Therefore, the energy required for the slip would thus be lower with dopants along the slip line, consistent with the overall doping induced decrement in GSFE shown in Figure 2a. We further plot the evolutions of the charge density and local bonding configuration in order to closely examine the slip process. Figure 3a shows the charge density plots for pristine graphene and graphene with dopants directly along the slip line at different stages during the slip process.[83] As the slip displacement $\delta$ increases, one of the σ bonds between C and C (or N, B) atoms (per unit cell) is gradually broken. Following the bond breakage, the atoms along the slip line will feature unpaired electrons. They can either interact with the adjacent atoms across the slip or with the neighboring π systems away from the slip. Figure 3b schematically illustrates the local bonding configurations at $\delta = 0.5b$ for pristine and impurity-doped graphene. In the cases of pristine and $N^{AS}$ doped graphene, C and N have more than one unpaired electrons, which can promote the formation of a triple bond across the slip to stabilize the system, resulting a local minimum, i.e., $\gamma_{sf}$, at $\delta = 0.5b$ (cf. Figure 2a). Nonetheless, the extra electron in the case of C-N pair likely leads to stronger charge interaction to further aid the triple bond formation, thus yielding a more stable bonding at $\delta = 0.5b$ (cf. Figure 2a). In the cases of B$^{AS}$ doping, since there is only one unpaired electron in B atom, triple bond formation across the slip is unlikely. Instead, the unpaired electron in B may pair with the neighboring π system away from the slip (cf. Figure 3b). Consequently no $\gamma_{sf}$ is expected in the case of B$^{AS}$ doping. The above bonding processes are also evidenced by the evolution of bond length[84] across the slip in Figure 3c, showing that the interatomic distance of C-N pair being the smallest while the one of C-B pair being the largest. Furthermore, we can note from the charge density plots in Figure 3a that (at $\delta = 0.5b$) there is strong presence of charge between C-N and C-C pairs but little charge density between the C-B pair across the slip.

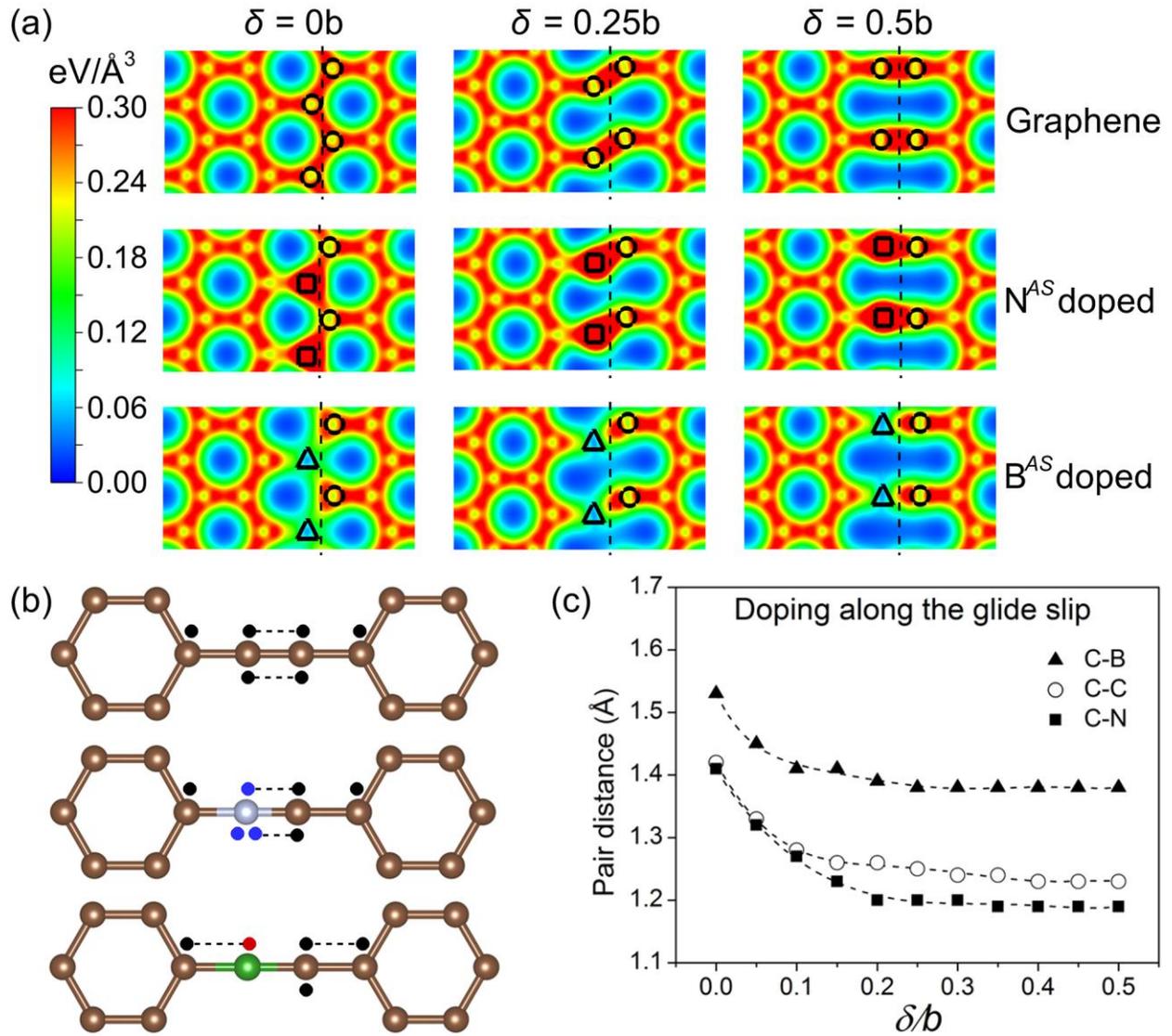

**Figure 3.** (a) Charge density plots for pristine graphene (top row), graphene with $N^{AS}$ doping (middle row) and graphene with $B^{AS}$ doping (bottom row) at three representative slip displacements along the glide line. Relevant C, N, and B atoms are highlighted by open circles, squares, and triangles respectively. (b) Schematic plot showing the bond reforming at $\delta = 0.5b$, where the relevant unpaired valence shell electrons of C, N, and B are indicated by black, blue and red dots respectively. The dotted lines represent the unpaired electron interactions. (c) The evolution of C-C, C-N, and C-B pair distance[85] across the slip line during the glide slip process.

When the dopants reside along the atomic row neighboring the slip line, they are not directly involved in the slip deformation and the bond breakage always occurs at the C-C bond. This is well indicated by the invariance in the GSFE curve at small slip deformation, i.e., $\delta <$ 0.15b (and symmetrically 0.85b $< \delta <$ b) regardless of dopant type or concentration (cf. Figure 2d). The influence of doping on the GSFE becomes noticeable at large slip deformation (i.e., 0.2b $< \delta <$ 0.8b) where one C-C bond (per unit cell) across the slip is broken. As previously mentioned, the unpaired electron resulted from bond breakage may either interact with the adjacent atoms across the slip or with the neighboring π systems, and the triple bond formation across the slip would lower the overall energy. From the charge density plots shown in Figure 4a (e.g., at $\delta = 0.5$b), $N^{NS}$ doping features high charge density between the C-N pair while in contrast $B^{NS}$ doping features very weak charge presence. Therefore the unpaired electron would favor interaction with its neighboring π system in the case of $N^{NS}$ doping but prefer formation of a triple bond in the case of $B^{NS}$ doping (cf. Figure 4b). The above competition associated with the unpaired electron is also well demonstrated by the evolution of bond length (i.e., for C-N, C-B and C-C bonds) shown in Figure 4c, where we note that beyond a small slip deformation the C-N and C-B pairs exhibit the shortest and largest bond lengths respectively. In particular we note that the dopant-induced bond length modification is much smaller in $N^{NS}$ doping than $B^{NS}$ doping, consistent with the magnitude of dopant-induced influence on the GSFE in Figure 2d.

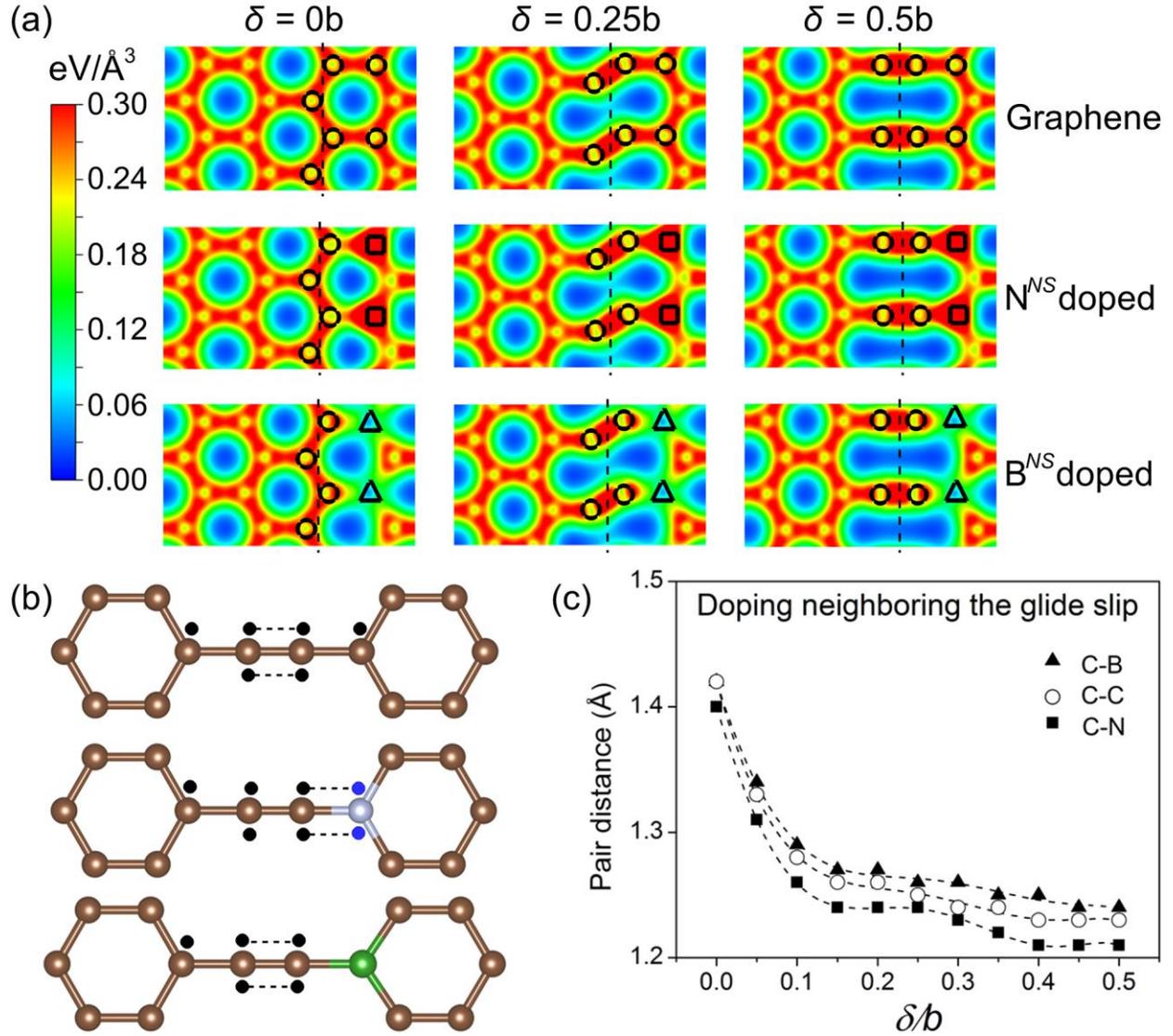

**Figure 4.** (a) Charge density plots for pristine graphene (top row), graphene with $N^{NS}$ doping (middle row) and graphene with $B^{NS}$ doping (bottom row) at three representative slip displacements along the glide line. Relevant C, N, and B atoms are highlighted by open circles, squares, and triangles respectively. (b) Schematic plot showing the bond reforming at $\delta = 0.5b$, where the relevant unpaired valence shell electrons of C and N are indicated by black and blue dots respectively. The dotted lines represent the unpaired electron interactions. (c) The evolution of C-C, C-N, and C-B pair distance[85] immediately neighboring the slip line during the glide slip process.

### 3.2. GSFE curves along shuffle direction

For the case of shuffle slip, the GSFE curves for the pristine graphene and two representative cases with $B^{AS}$ and $N^{AS}$ doping along the slip line respectively are presented in Figure 5a, showing that doping lowers the GSFE curve. It is shown in Figure 5b that $\gamma_{usf}$ decreases roughly in a linear fashion as the dopant ($B^{AS}$ and $N^{AS}$) concentration increases, with the reduction in $\gamma_{usf}$ being more pronounced in the case of $N^{AS}$ doping. One thing to note is that unlike the GSFE curve for the glide slip, the GSFE curve for the shuffle slip does not exhibit a meta-stable stacking fault, with or without dopants (either along or neighboring the slip line). Thus $\gamma_{sf}$ is non-existent for the shuffle slip. Another observation worth noting is that $N^{AS}$ doping leads to a seemingly wide plateau in the middle of GSFE curve (i.e., the regime between $\delta = 0.25b$ and $\delta = 0.75b$). On the other hand, the influence of dopants on the GSFE is rather limited when they sit along the atomic row immediately neighboring the shuffle slip line, shown in Figures 5c-d. From the plots of $\gamma_{usf}$ as functions of dopant concentration in Figure 5d, we note that $B^{NS}$ doping hardly affect $\gamma_{usf}$, while $N^{NS}$ doping leads to small but noticeable reduction in $\gamma_{usf}$. Additionally, $N^{NS}$ doping also tends to flatten the GSFE curve around $\delta = 0.5b$ (cf. Figure 5c), similar to the case of $N^{AS}$ doping (cf. Figure 5a).

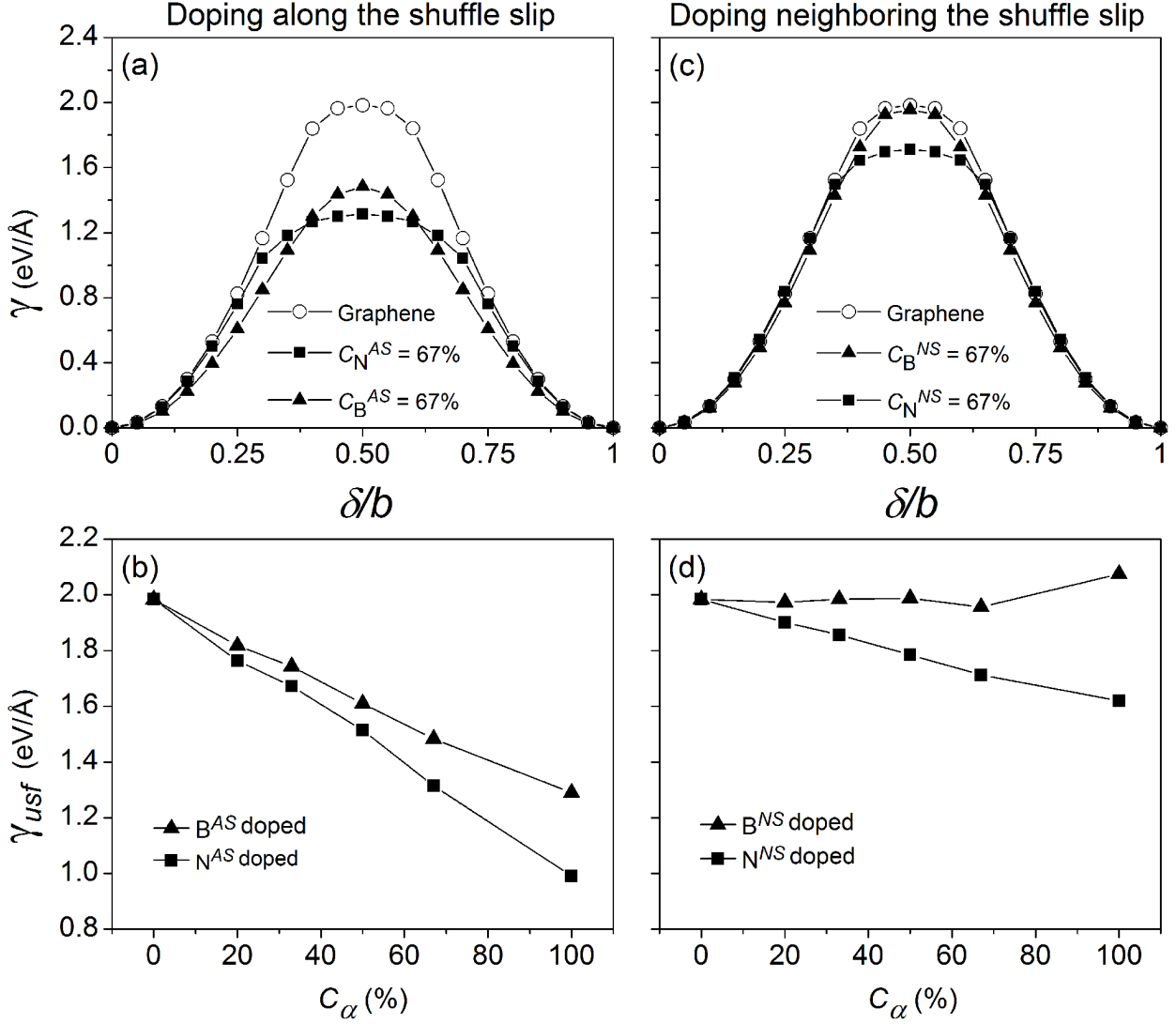

**Figure 5.** Sample GSFE curves for graphene (open circles) and graphene with B (solid triangles) and N (solid squares) doping along (a) the shuffle slip line and (c) atomic row immediately neighboring the shuffle slip line. The unstable $\gamma_{usf}$ as functions of the dopant concentration is shown in (b) for doping along the shuffle slip line, and (d) for doping along atomic row immediately neighboring the shuffle slip line. The acronyms *AS* and *NS* indicate dopants along and immediately neighboring the slip line respectively.

The corresponding charge density contours for the shuffle slip is illustrated in Figures 6a and 6b for the pristine graphene, graphene with $N^{AS}$ and $N^{NS}$ doping and graphene with $B^{AS}$ and $B^{NS}$ doping respectively. Like the case of the glide slip with impurities along the slip line, doping renders some bonds across the slip plane from the stronger C-C to weaker C-N and C-B bonds, thus reducing the net energy required for breaking the bonds during the slip and lowering the GSFE (cf. Figure 5a). As seen in Figure 6a, the charge density along the slip line reaches the minimum level at $\delta = 0.5b$, showing continuous wavy pathway with virtually zero charge. This suggests that no meta-stable stacking fault exists in shuffle slip and at $\delta = 0.5b$ where $\gamma_{usf}$ ensues, all bonds across the shuffle slip line are broken. When the dopants reside along the atomic row neighboring the shuffle slip line, their influence on the GSFE is much limited as expected. Similar to those shown in Figure 6a, the charge density contours presented in Figure 6b also clearly show a near zero charge density pathway following the breaking of C-C bonds across the slip, suggesting the absence of $\gamma_{sf}$. Another observation drawn from the charge density contours in Figure 6 is that the strong charge presence associated with N atoms seemingly helps strengthen C-N interactions post the bond breakage. This compensates the energy cost required for slipping, providing a possible explanation for the middle flat plateau in the GSFE curve under $N^{AS}$ (cf. Figure 5a) and $N^{NS}$ (cf. Figure 5c) doping.

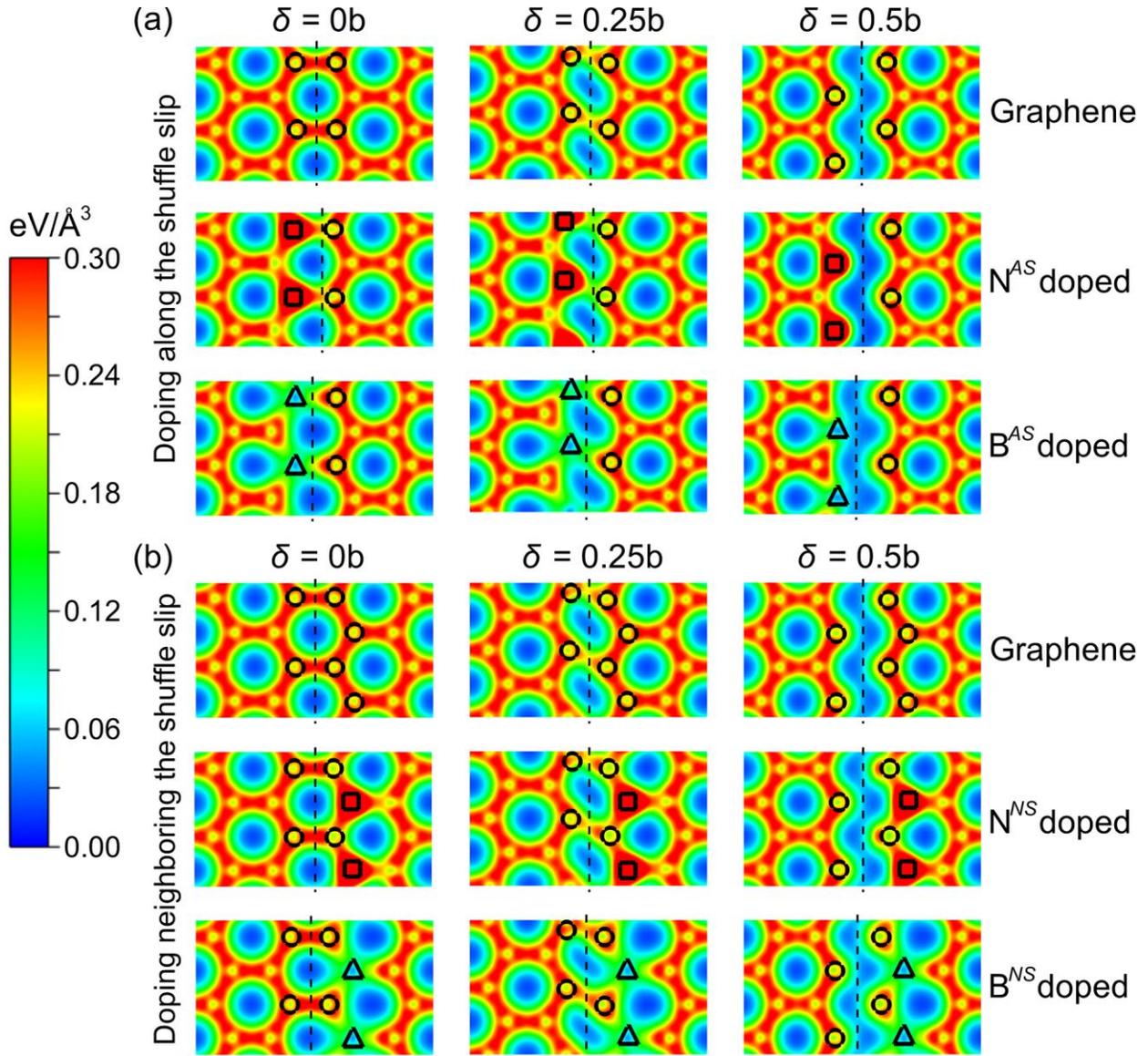

**Figure 6.** Charge density plots for (a) pristine graphene (top row), graphene with $N^{AS}$ doping (middle row) and graphene with $B^{AS}$ doping (bottom row) and (b) pristine graphene (top row), graphene with $N^{NS}$ doping (middle row) and graphene with $B^{NS}$ doping (bottom row) at three representative slip displacements along the shuffle line. Relevant C, N, and B atoms are highlighted by open circles, squares, and triangles respectively.

### 3.3. Effects of doping on dislocation dipoles and core dissociation

The GSFE curves provide essential information on the stability of dislocation dipoles and structural characteristics of single dislocations. Considering a dislocation dipole where the two dislocations are separated by a distance $d$, it would be stable against annealing under the following condition:

$$d \geq d_{pa} = \frac{Kb^2}{4\pi\gamma_{usf}}, \qquad (2)$$

where $d_{pa}$ is the minimum spacing to avoid annihilation, i.e., the *spontaneous pair-annihilation* distance, and $K$ is the effective elastic constant.[74] The value of $K$ was obtained by Ariza and Ortiz as 15.49 eV/Å$^2$. Equation 2 above derives from the stability condition that the elastic energy released by annihilation cannot exceed the energy barrier $\gamma_{usf}$. The $d_{pa}$ values as functions of dopant concentration $c_\alpha$ are shown in Figures. 7a and 7b for the glide and shuffle slips respectively.

For the glide slip, the pristine graphene exhibits a $d_{pa}$ of 2.1b, close to the value previously reported by Ariza et al.[74] using an AIREBO potential. From Figure 7a, we see that $d_{pa}$ increases under $B^{AS}$, $N^{AS}$ and $B^{NS}$ doping while slightly decreases under $N^{NS}$ doping. For the shuffle slip, it is shown in Figure 7b that $d_{pa}$ increases under $B^{AS}$, $N^{AS}$ and $N^{NS}$ doping while remains largely unchanged under $B^{NS}$ doping. In both the glide and shuffle slip, we see that doping can lead to as much as one-fold increase in $d_{pa}$. This suggests that B or N doping (particularly along the slip) provides a way to eliminate dislocation dipoles of small separations and thus reduce dislocation density in graphene. However, given that $d_{pa}$ remains quite small (< 5b) even at very high dopant concentrations (cf. Figures 7a-b), the doping-induced dipole annihilation would only become relevant in heavily deformed graphene.

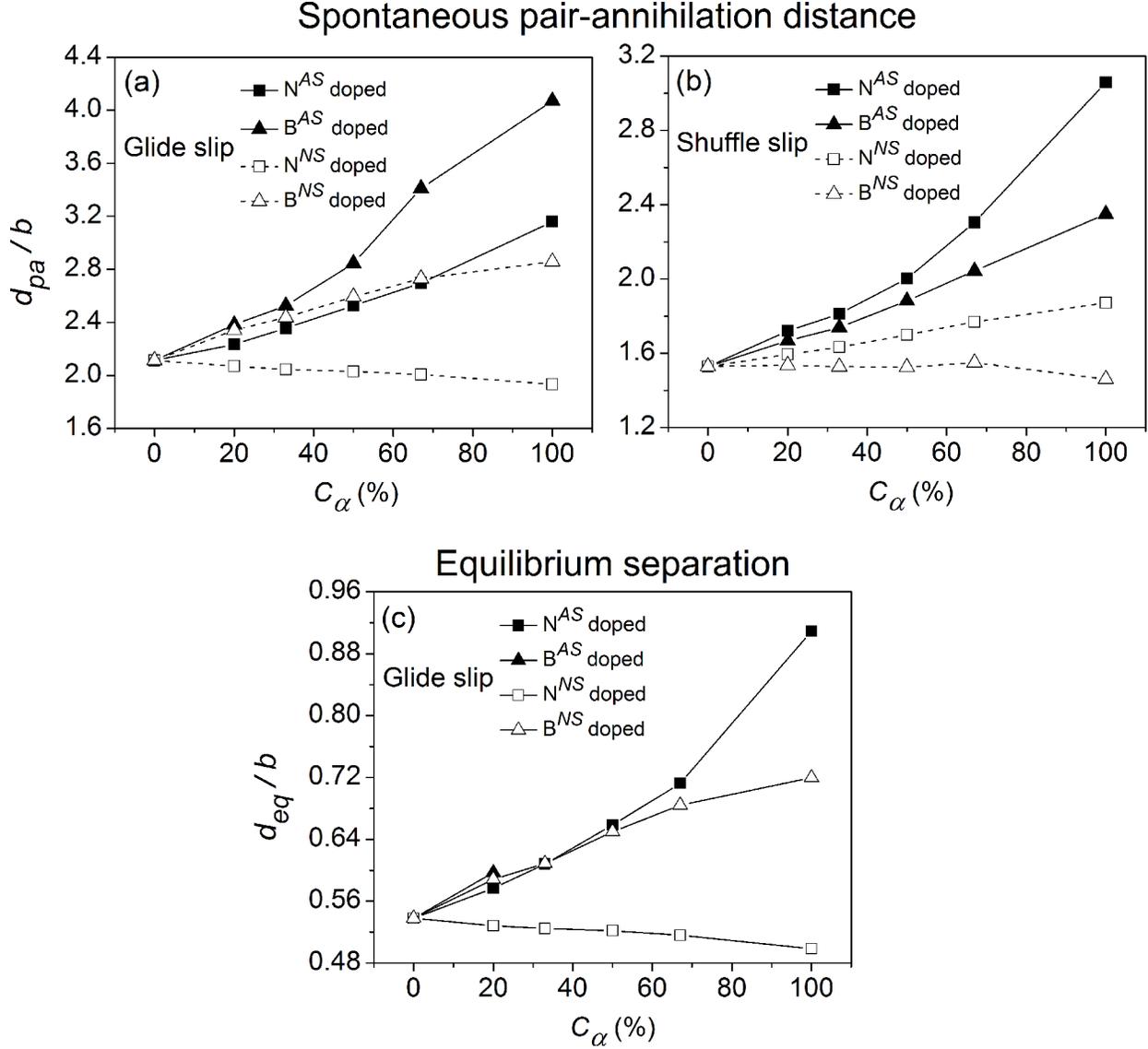

**Figure 7.** The spontaneous pair-annihilation distance, $d_{pa}$, normalized by the magnitude of the Burgers vector, $b$, for the (a) glide slip and (b) shuffle slip, as functions of dopant concentrations. (c) The equilibrium separation, $d_{eq}$, between two split dislocation partials in the glide slip. The solid square, solid triangle, open square and open triangle indicate $N^{AS}$, $B^{AS}$ (no $\gamma_{sf}$ exists for doping $B^{AS}$ and thus no $d_{eq}$ is calculated for $c_B > 20\%$), $N^{NS}$ and $B^{NS}$ doping respectively.

For a standing-alone dislocation, the GSFE enables quantitative analysis of the core structure. Considering a simple scenario where the dislocation core is symmetrically split into two partials of Burgers vector $b/2$ separated by a stacking-fault ribbon. Balancing the repulsive elastic force and the attracted force due to the stacking-fault ribbon leads to:[86]

$$\frac{K(b/2)^2}{4\pi d} = \gamma_{sf}, \tag{3}$$

which gives the equilibrium separation, $d_{eq}$, between the partials:

$$d_{eq} = \frac{Kb^2}{16\pi\gamma_{sf}}. \tag{4}$$

The core dissociation is only relevant in the case of glide slip as the shuffle slip does not exhibits a meta-stable stacking fault in the GSFE. The $d_{eq}$ values as functions of dopant concentration $c_\alpha$ are shown in Figure 7c (no $\gamma_{sf}$ exists for doping $B^{AS}$ from $c_B > 20\%$), from which we can see that the $N^{AS}$ and $B^{NS}$ doping promote core splitting while $N^{NS}$ inhibits core splitting. In the case of $B^{AS}$ doping, it favors core splitting at small dopant concentration, i.e., $c_B < 20\%$, but completely forbids core splitting at higher dopant concentration. Nonetheless overall we see from Figure 7c that $d_{eq}$ is always less than b,[87] implying in general core dissociation does not happen.

### 3.4. Micromechanical analysis using Peierls-Nabarro (P-N) model

The GSFE curve also provides essential inputs for the P-N model[61-62] to enable analysis of dislocation motions within the continuum framework. The motions of dislocations are directly responsible for plastic deformation in the material. The slip of a straight dislocation through the crystal lattice results in periodic variation in energy, derived from misfit energy $W(u)$ over the slip plane. For a narrow dislocation, $W(u)$ is approximated as:[65]

$$W(u) = \frac{Kb^2 a'}{4\pi^2} \frac{\xi}{\xi^2 + u^2}, \tag{5}$$

where $K$ is the effective elastic constant previously introduced (cf. eq 2), $a'$ is the atomic repeat distance along the slip direction, set as b for graphene,[88] $\xi$ is the half width or core radius of the dislocation, and $u$ is the dislocation translational distance. To enable the dislocation motion, an energy barrier, i.e., the Peierls barrier, defined as the maximal variation of $W(u)$, has to be overcome.[86] The minimum stress to drive the dislocation over the Peierls barrier $W_p$ is the Peierls stress $\sigma_p$.[61-62, 86, 89] $\sigma_p$ is defined as the maximum derivative of the misfit energy:[65]

$$\sigma_p = \max[\sigma(u)] = \max\left[\frac{1}{b}\frac{dW(u)}{du}\right] = \max\left[-\frac{Kba'}{2\pi^2}\frac{\xi u}{(\xi^2+u^2)^2}\right], \quad (6)$$

which follows to yield:[65]

$$\sigma_P = \frac{3\sqrt{3}}{8}\tau_{max}\frac{a'}{\pi\xi}, \quad (7)$$

where $\tau_{max}$ is the maximal slope of the corresponding GSFE curve,[90] and can be regarded as the theoretical shear strength along the slip direction. The parameter $\xi$ is related to $\tau_{max}$ as:[65]

$$\xi = \frac{Kb}{4\pi\tau_{max}}. \quad (8)$$

Equations 7-8 were shown to provide good predictions of the Peierls stress for narrow dislocations satisfying $\xi/a' < 1$.[65]

Using the above eqs 5-8, we evaluate $W_p$ (Peierls barrier), $\sigma_p$ (Peierls stress), $\tau_{max}$ (theoretical shear strength) and $\xi$ (dislocation core radius) for dislocations in pristine and impurity (B or N) doped graphene, shown in Figures 8 and 9 for the glide and shuffle slip respectively. For pristine graphene, the dislocation core radius is calculated to be 0.91 Å for the glide slip and 1.08 Å for the shuffle slip, consistent with core radii reported, e.g., 1.20 Å by fitting local-density approximation calculations[91] and 0.96 Å by using least-squares fit of the Read-Shockley equation.[44] Also we can note that in pristine graphene the glide direction is more resistant to slip, exhibiting higher $W_p$, $\sigma_p$, and $\tau_{max}$ than the shuffle direction. The higher slip resistance may derive from the fact that the glide slip involves distortion/breakage of two bonds per atom compared to just one bond per atom during the shuffle slip, as suggested in Ref. 74.

For impurity-doped graphene, the results for the glide slip are shown in Figure 8, showing that $W_p$, $\sigma_p$ and $\tau_{max}$ decrease while $\zeta$ increases with increasing amount of $B^{AS}$, $N^{AS}$ or $B^{NS}$ doping. The opposite trend in $\zeta$ with respect to the other three parameters is expected as a wider core would lead to larger Peierls barrier and Peierls stress.[63, 92] On the other hand, the $N^{NS}$ doping exerts little influence on all four parameters. In the case of the shuffle slip, we see from Figure 9 that $W_p$, $\sigma_p$ and $\tau_{max}$ decrease while $\zeta$ increases with increasing amount of $B^{AS}$ or $N^{AS}$ doping. Meanwhile these parameters are rather indifferent to $B^{NS}$ and $N^{NS}$ doping. Also we can note that overall the effect of doping is less pronounced for the shuffle slip than the glide slip, particularly for N. In particular we see that with sufficient $B^{AS}$ doping, $W_p$, $\sigma_p$ and $\tau_{max}$ of the glide slip can be rendered to be much lower than the shuffle slip despite the glide direction being more slip-resistant in pristine graphene.

The shear strength of graphene was previously studied by Min and Aluru,[93] being ~ 60 GPa. From Figures 8 and 9, the Peierls stresses are 88 GPa and 63 GPa for the glide and shuffle slips respectively, both being higher than those shear strength values reported, suggesting that dislocation motions are not possible in pristine graphene. Nonetheless, the introduction of B or N dopants can reduce the Peierls stress below the shear strength of graphene (cf. Figures 8d and 9d) to activate dislocation motions, and thus facilitate plasticity in graphene.

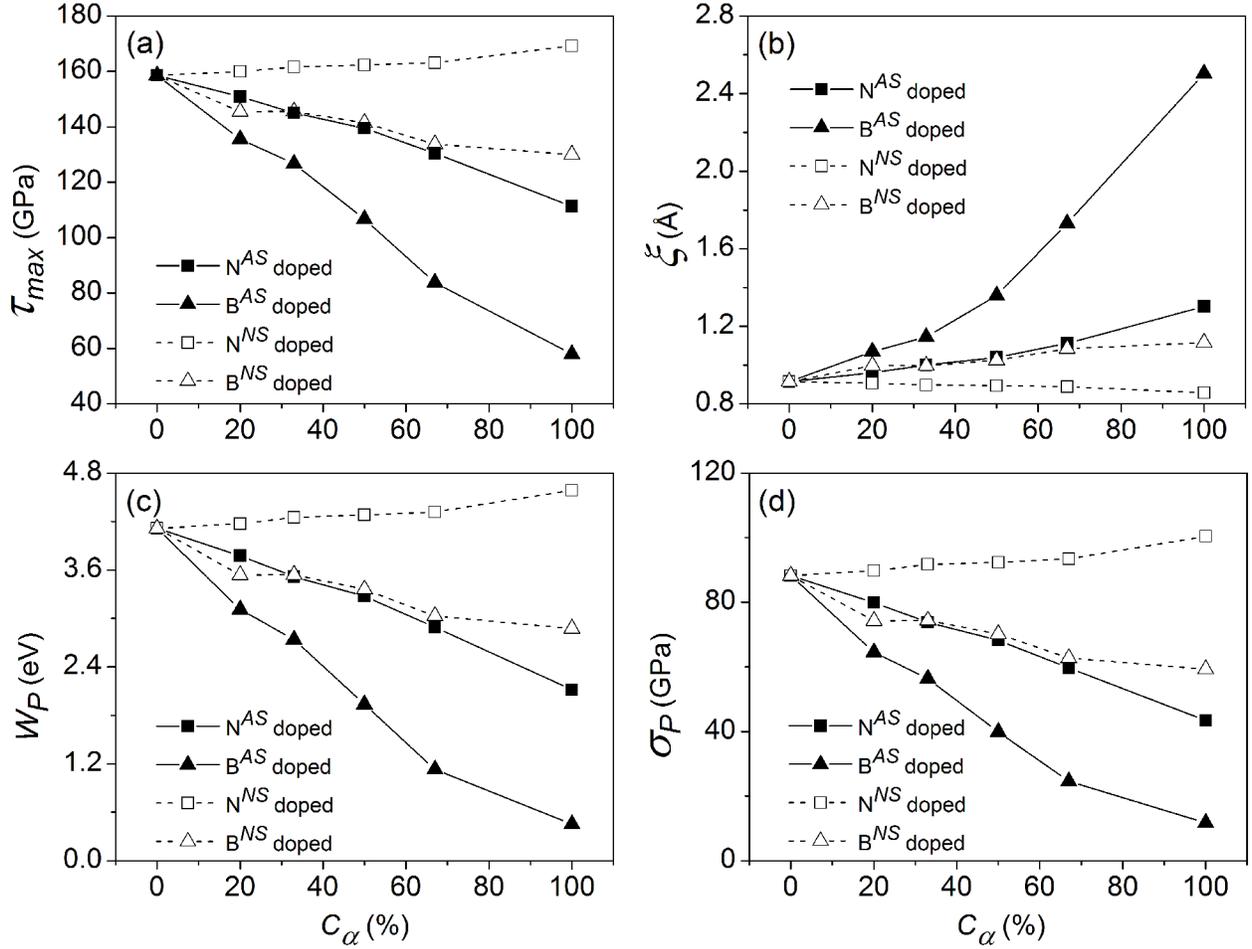

**Figure 8.** The evolution of (a) theoretical shear strength, $\tau_{max}$, (b) core radius, $\xi$, (c) Peierls barrier, $W_p$, and (d) Peierls stress, $\sigma_p$, as the dopant concentration varies. The solid square, solid triangle, open square and open triangle indicate $N^{AS}$, $B^{AS}$, $N^{NS}$ and $B^{NS}$ doping respectively.

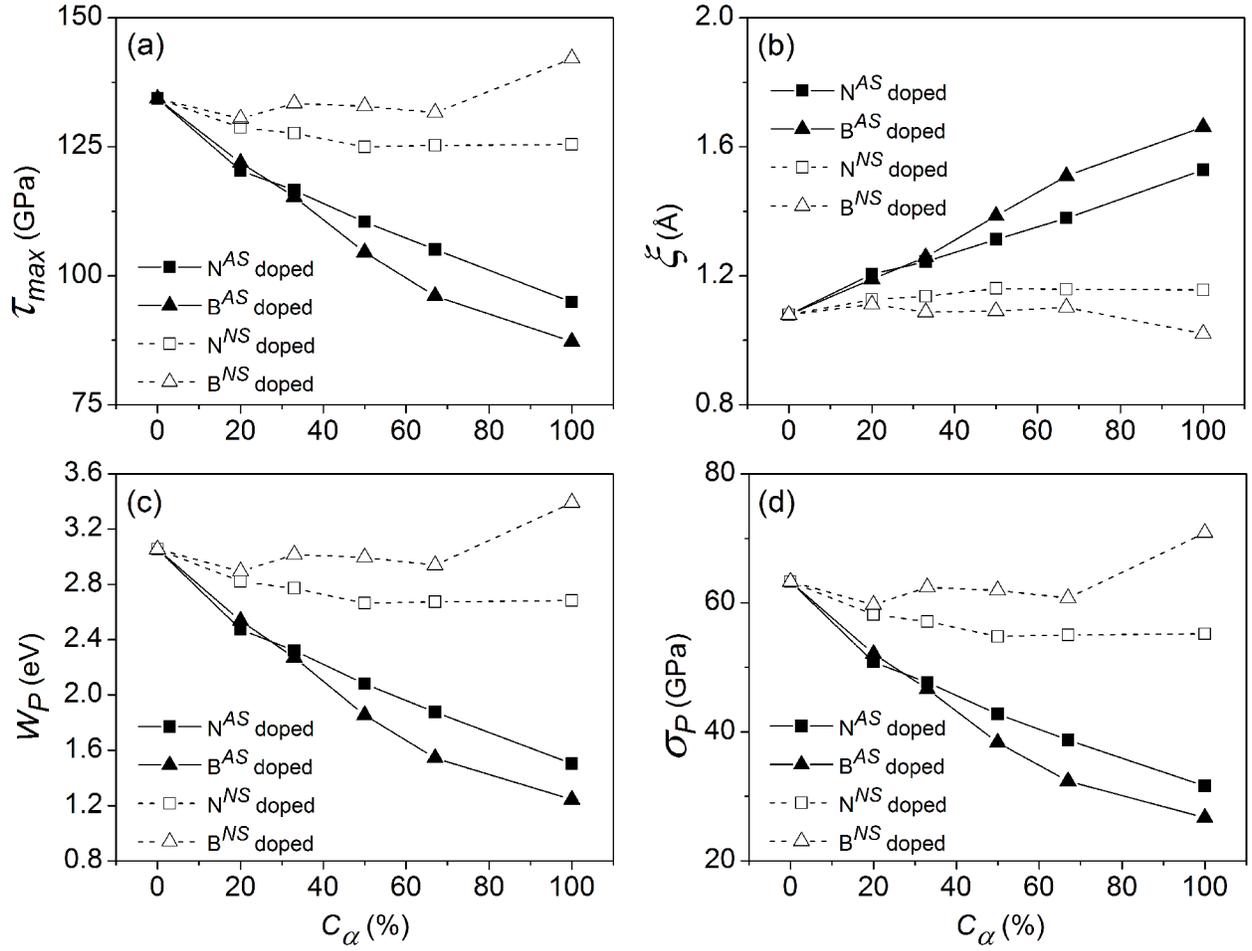

**Figure 9.** The evolution of (a) theoretical shear strength, $\tau_{max}$, (b) core radius, $\xi$, (c) Peierls barrier, $W_p$, and (d) Peierls stress, $\sigma_p$, as the dopant concentration varies. The solid square, solid triangle, open square and open triangle indicate $N^{AS}$, $B^{AS}$, $N^{NS}$ and $B^{NS}$ doping respectively.

## IV. CONCLUSIONS

In summary, the generalized-stacking-fault energy (GSFE) curves along two crystallographic slips, glide and shuffle, for both pristine graphene and impurity (i.e., B or N) doped graphene were computed using density-functional theory (DFT) calculations. B or N dopants of different concentrations were introduced into graphene either along or immediately neighboring the slip line. For the glide slip, the incorporation of substitutional B or N atoms along the slip line was shown to overall lower the GSFE curve, which is expected given the positive formation energies of B and N in graphene. Meanwhile the shape of the GSFE curve is also altered due to the presence of those dopants. In particular, N dopants along the slip line help stabilize the meta-stable stacking fault, attributed to their extra valence electrons that facilitate the formation of a triple bond, while on the contrary the meta-stable stacking fault is moderated or eliminated under B doping along the slip due to insufficient unpaired electrons for triple bond formation. In cases where dopants reside immediately neighboring the glide slip line, the GSFE curve is retained albeit the magnitude of GSFE changes due to the modified electron interactions with C-C bonding along the slip. On the other hand the shuffle slip does not exhibit a meta-stable stacking fault regardless of the presence of dopants. Similar to the glide slip, the presence of along-the-slip dopants overall lowers the GSFE curve, and dopants immediately neighboring the shuffle slip yield limited influence on the GSFE curve. One particular observation concerning the shuffle slip is that N doping (either along or immediately neighboring the slip line) leads to a flat plateau amidst the GSFE curve. In addition, one thing worth noting from our results is that in general the effects of doping on the GSFE are closely related to the dopant induced modification of the local charge density. This hints the possibility of engineering of the GSFE and subsequently dislocation slip through charge injection.

Based on the GSFE data, we showed that doping (particularly along the slip) can increase the spontaneous pair-annihilation distance for dislocation dipoles, providing a means to reduce dislocation densities in heavily deformed graphene. We also demonstrated that doping may affect the equilibrium splitting distance between dislocation partials, but not sufficient to drive actual core dissociation. The GSFE data then meshed with the Peierls-Nabarro (P-N) model to analyze dislocations within the continuum framework. The slip deformation was shown to be considerably facilitated with along-the-slip doing of B and N, but rather indifferent to doping neighboring the slip line. The present findings provide fundamental information on the effects of B and N doping on microscopic plasticity in graphene, and can be used to aid defect engineering in graphene-based materials.

**ACKNOWLEDGEMENT**

We greatly thank the financial support from McGill Engineering Doctoral Award and National Sciences and Engineering Research Council (NSERC) Discovery grant (grant # RGPIN 418469-2012). We also acknowledge Supercomputer Consortium Laval UQAM McGill and Eastern Quebec for providing computing power.

**SUPPORTING INFORMATION AVAILABLE**

Sample generalized-stacking-fault curves show the combinative effects of boron and nitrogen dopants on boosting the equilibrium distance between dislocation partials and reducing the Peierls stress for dislocation motion in graphene. This information is available free of charge via the Internet at http://pubs.acs.org.

# REFERENCES

1. Novoselov, K. S.; Geim, A. K.; Morozov, S.; Jiang, D.; Zhang, Y.; Dubonos, S.; Grigorieva, I.; Firsov, A., Electric Field Effect in Atomically Thin Carbon Films. *Science* **2004**, *306*, 666-669.
2. Frank, I.; Tanenbaum, D. M.; Van der Zande, A.; McEuen, P. L., Mechanical Properties of Suspended Graphene Sheets. *J. Vac. Sci. Technol., B* **2007**, *25*, 2558-2561.
3. Geim, A. K.; Novoselov, K. S., The Rise of Graphene. *Nat. Mater.* **2007**, *6*, 183-191.
4. Balandin, A. A.; Ghosh, S.; Bao, W.; Calizo, I.; Teweldebrhan, D.; Miao, F.; Lau, C. N., Superior Thermal Conductivity of Single-Layer Graphene. *Nano Lett.* **2008**, *8*, 902-907.
5. Bolotin, K. I.; Sikes, K.; Jiang, Z.; Klima, M.; Fudenberg, G.; Hone, J.; Kim, P.; Stormer, H., Ultrahigh Electron Mobility in Suspended Graphene. *Solid State Commun.* **2008**, *146*, 351-355.
6. Neto, A. C.; Guinea, F.; Peres, N.; Novoselov, K. S.; Geim, A. K., The Electronic Properties of Graphene. *Rev. Mod. Phys.* **2009**, *81*, 109-162.
7. Novoselov, K.; Geim, A. K.; Morozov, S.; Jiang, D.; Grigorieva, M. K. I.; Dubonos, S.; Firsov, A., Two-Dimensional Gas of Massless Dirac Fermions in Graphene. *Nature* **2005**, *438*, 197-200.
8. Zhang, Y.; Tan, Y.-W.; Stormer, H. L.; Kim, P., Experimental Observation of the Quantum Hall Effect and Berry's Phase in Graphene. *Nature* **2005**, *438*, 201-204.
9. Berger, C.; Song, Z.; Li, X.; Wu, X.; Brown, N.; Naud, C.; Mayou, D.; Li, T.; Hass, J.; Marchenkov, A. N., Electronic Confinement and Coherence in Patterned Epitaxial Graphene. *Science* **2006**, *312*, 1191-1196.
10. Ferrari, A.; Meyer, J.; Scardaci, V.; Casiraghi, C.; Lazzeri, M.; Mauri, F.; Piscanec, S.; Jiang, D.; Novoselov, K.; Roth, S., Raman Spectrum of Graphene and Graphene Layers. *Phys. Rev. Lett.* **2006**, *97*, 187401 1-4.
11. Han, M. Y.; Özyilmaz, B.; Zhang, Y.; Kim, P., Energy Band-Gap Engineering of Graphene Nanoribbons. *Phys. Rev. Lett.* **2007**, *98*, 206805 1-4.
12. Nair, R.; Blake, P.; Grigorenko, A.; Novoselov, K.; Booth, T.; Stauber, T.; Peres, N.; Geim, A., Fine Structure Constant Defines Visual Transparency of Graphene. *Science* **2008**, *320*, 1308-1308.
13. Lin, Y.-M.; Dimitrakopoulos, C.; Jenkins, K. A.; Farmer, D. B.; Chiu, H.-Y.; Grill, A.; Avouris, P., 100-Ghz Transistors from Wafer-Scale Epitaxial Graphene. *Science* **2010**, *327*, 662-662.
14. Stankovich, S.; Dikin, D. A.; Dommett, G. H.; Kohlhaas, K. M.; Zimney, E. J.; Stach, E. A.; Piner, R. D.; Nguyen, S. T.; Ruoff, R. S., Graphene-Based Composite Materials. *Nature* **2006**, *442*, 282-286.
15. Rafiee, M. A.; Rafiee, J.; Wang, Z.; Song, H.; Yu, Z.-Z.; Koratkar, N., Enhanced Mechanical Properties of Nanocomposites at Low Graphene Content. *ACS nano* **2009**, *3*, 3884-3890.
16. Zandiatashbar, A.; Picu, C. R.; Koratkar, N., Control of Epoxy Creep Using Graphene. *Small* **2012**, *8*, 1676-1682.
17. Bunch, J. S.; Verbridge, S. S.; Alden, J. S.; van der Zande, A. M.; Parpia, J. M.; Craighead, H. G.; McEuen, P. L., Impermeable Atomic Membranes from Graphene Sheets. *Nano Lett.* **2008**, *8*, 2458-2462.

18. Nair, R.; Wu, H.; Jayaram, P.; Grigorieva, I.; Geim, A., Unimpeded Permeation of Water through Helium-Leak–Tight Graphene-Based Membranes. *Science* **2012**, *335*, 442-444.
19. Schedin, F.; Geim, A.; Morozov, S.; Hill, E.; Blake, P.; Katsnelson, M.; Novoselov, K., Detection of Individual Gas Molecules Adsorbed on Graphene. *Nat. Mater.* **2007**, *6*, 652-655.
20. Yoon, H. J.; Jun, D. H.; Yang, J. H.; Zhou, Z.; Yang, S. S.; Cheng, M. M.-C., Carbon Dioxide Gas Sensor Using a Graphene Sheet. *Sensors Actuators B: Chem.* **2011**, *157*, 310-313.
21. Paul, R. K.; Badhulika, S.; Saucedo, N. M.; Mulchandani, A., Graphene Nanomesh as Highly Sensitive Chemiresistor Gas Sensor. *Anal. Chem.* **2012**, *84*, 8171-8178.
22. Yavari, F.; Castillo, E.; Gullapalli, H.; Ajayan, P. M.; Koratkar, N., High Sensitivity Detection of No2 and Nh3 in Air Using Chemical Vapor Deposition Grown Graphene. *Appl. Phys. Lett.* **2012**, *100*, 203120 1-4.
23. Hashimoto, A.; Suenaga, K.; Gloter, A.; Urita, K.; Iijima, S., Direct Evidence for Atomic Defects in Graphene Layers. *Nature* **2004**, *430*, 870-873.
24. Rutter, G.; Crain, J.; Guisinger, N.; Li, T.; First, P.; Stroscio, J., Scattering and Interference in Epitaxial Graphene. *Science* **2007**, *317*, 219-222.
25. Meyer, J. C.; Kisielowski, C.; Erni, R.; Rossell, M. D.; Crommie, M.; Zettl, A., Direct Imaging of Lattice Atoms and Topological Defects in Graphene Membranes. *Nano Lett.* **2008**, *8*, 3582-3586.
26. Robertson, A. W.; Warner, J. H., Atomic Resolution Imaging of Graphene by Transmission Electron Microscopy. *Nanoscale* **2013**, *5*, 4079-4093.
27. Stone, A. J.; Wales, D. J., Theoretical Studies of Icosahedral C60 and Some Related Species. *Chem. Phys. Lett.* **1986**, *128*, 501-503.
28. Coraux, J.; N'Diaye, A. T.; Busse, C.; Michely, T., Structural Coherency of Graphene on Ir (111). *Nano Lett.* **2008**, *8*, 565-570.
29. Park, H. J.; Meyer, J.; Roth, S.; Skákalová, V., Growth and Properties of Few-Layer Graphene Prepared by Chemical Vapor Deposition. *Carbon* **2010**, *48*, 1088-1094.
30. Kim, K.; Lee, Z.; Regan, W.; Kisielowski, C.; Crommie, M.; Zettl, A., Grain Boundary Mapping in Polycrystalline Graphene. *ACS nano* **2011**, *5*, 2142-2146.
31. Huang, P. Y.; Ruiz-Vargas, C. S.; van der Zande, A. M.; Whitney, W. S.; Levendorf, M. P.; Kevek, J. W.; Garg, S.; Alden, J. S.; Hustedt, C. J.; Zhu, Y., Grains and Grain Boundaries in Single-Layer Graphene Atomic Patchwork Quilts. *Nature* **2011**, *469*, 389-392.
32. Reina, A.; Jia, X.; Ho, J.; Nezich, D.; Son, H.; Bulovic, V.; Dresselhaus, M. S.; Kong, J., Large Area, Few-Layer Graphene Films on Arbitrary Substrates by Chemical Vapor Deposition. *Nano Lett.* **2008**, *9*, 30-35.
33. Li, X.; Cai, W.; An, J.; Kim, S.; Nah, J.; Yang, D.; Piner, R.; Velamakanni, A.; Jung, I.; Tutuc, E., Large-Area Synthesis of High-Quality and Uniform Graphene Films on Copper Foils. *Science* **2009**, *324*, 1312-1314.
34. Gao, L.; Guest, J. R.; Guisinger, N. P., Epitaxial Graphene on Cu (111). *Nano Lett.* **2010**, *10*, 3512-3516.
35. Khare, R.; Mielke, S. L.; Paci, J. T.; Zhang, S.; Ballarini, R.; Schatz, G. C.; Belytschko, T., Coupled Quantum Mechanical/Molecular Mechanical Modeling of the Fracture of Defective Carbon Nanotubes and Graphene Sheets. *Phys. Rev. B* **2007**, *75*, 075412 1-12.
36. Wang, S.-P.; Guo, J.-G.; Zhou, L.-J., Influence of Stone–Wales Defects on Elastic Properties of Graphene Nanofilms. *Physica E: Low-dimensional Systems and Nanostructures* **2013**, *48*, 29-35.

37. Grantab, R.; Shenoy, V. B.; Ruoff, R. S., Anomalous Strength Characteristics of Tilt Grain Boundaries in Graphene. *Science* **2010**, *330*, 946-948.
38. Liu, T.-H.; Pao, C.-W.; Chang, C.-C., Effects of Dislocation Densities and Distributions on Graphene Grain Boundary Failure Strengths from Atomistic Simulations. *Carbon* **2012**, *50*, 3465-3472.
39. Zhang, J.; Zhao, J.; Lu, J., Intrinsic Strength and Failure Behaviors of Graphene Grain Boundaries. *ACS nano* **2012**, *6*, 2704-2711.
40. Kotakoski, J.; Meyer, J. C., Mechanical Properties of Polycrystalline Graphene Based on a Realistic Atomistic Model. *Phys. Rev. B* **2012**, *85*, 195447 1-6.
41. Wei, Y.; Wu, J.; Yin, H.; Shi, X.; Yang, R.; Dresselhaus, M., The Nature of Strength Enhancement and Weakening by Pentagon–Heptagon Defects in Graphene. *Nat. Mater.* **2012**, *11*, 759-763.
42. Han, J.; Ryu, S.; Sohn, D.; Im, S., Mechanical Strength Characteristics of Asymmetric Tilt Grain Boundaries in Graphene. *Carbon* **2014**, *68*, 250-257.
43. Zandiatashbar, A.; Lee, G.-H.; An, S. J.; Lee, S.; Mathew, N.; Terrones, M.; Hayashi, T.; Picu, C. R.; Hone, J.; Koratkar, N., Effect of Defects on the Intrinsic Strength and Stiffness of Graphene. *Nat. Commun.* **2014**, *5*, 1-9.
44. Yazyev, O. V.; Louie, S. G., Topological Defects in Graphene: Dislocations and Grain Boundaries. *Phys. Rev. B* **2010**, *81*, 195420 1-7.
45. Carpio, A.; Bonilla, L., Periodized Discrete Elasticity Models for Defects in Graphene. *Phys. Rev. B* **2008**, *78*, 085406 1-11.
46. Ariza, M.; Ortiz, M.; Serrano, R., Long-Term Dynamic Stability of Discrete Dislocations in Graphene at Finite Temperature. *Int. J. Fract.* **2010**, *166*, 215-223.
47. Dean, C.; Young, A.; Meric, I.; Lee, C.; Wang, L.; Sorgenfrei, S.; Watanabe, K.; Taniguchi, T.; Kim, P.; Shepard, K., Boron Nitride Substrates for High-Quality Graphene Electronics. *Nat. Nanotechnol.* **2010**, *5*, 722-726.
48. Boukhvalov, D.; Katsnelson, M., Chemical Functionalization of Graphene with Defects. *Nano Lett.* **2008**, *8*, 4373-4379.
49. Gan, Y.; Sun, L.; Banhart, F., One‐ and Two‐Dimensional Diffusion of Metal Atoms in Graphene. *Small* **2008**, *4*, 587-591.
50. Krasheninnikov, A.; Lehtinen, P.; Foster, A.; Pyykkö, P.; Nieminen, R., Embedding Transition-Metal Atoms in Graphene: Structure, Bonding, and Magnetism. *Phys. Rev. Lett.* **2009**, *102*, 126807 1-4.
51. Santos, E. J.; Ayuela, A.; Sánchez-Portal, D., First-Principles Study of Substitutional Metal Impurities in Graphene: Structural, Electronic and Magnetic Properties. *New J. Phys.* **2010**, *12*, 053012 1-32.
52. Ci, L.; Song, L.; Jin, C.; Jariwala, D.; Wu, D.; Li, Y.; Srivastava, A.; Wang, Z.; Storr, K.; Balicas, L., Atomic Layers of Hybridized Boron Nitride and Graphene Domains. *Nat. Mater.* **2010**, *9*, 430-435.
53. Banhart, F.; Kotakoski, J.; Krasheninnikov, A. V., Structural Defects in Graphene. *ACS nano* **2010**, *5*, 26-41.
54. Ouyang, B.; Meng, F.; Song, J., Energetics and Kinetics of Vacancies in Monolayer Graphene Boron Nitride Heterostructures. *2D Mater.* **2014**, *1*, 035007 1-17.
55. Czerw, R.; Terrones, M.; Charlier, J.-C.; Blase, X.; Foley, B.; Kamalakaran, R.; Grobert, N.; Terrones, H.; Tekleab, D.; Ajayan, P., Identification of Electron Donor States in N-Doped Carbon Nanotubes. *Nano Lett.* **2001**, *1*, 457-460.

56. Terrones, H.; Terrones, M.; Hernández, E.; Grobert, N.; Charlier, J.-C.; Ajayan, P., New Metallic Allotropes of Planar and Tubular Carbon. *Phys. Rev. Lett.* **2000**, *84*, 1716 1-4.
57. Xiao, K.; Liu, Y.; Hu, P. a.; Yu, G.; Sun, Y.; Zhu, D., N-Type Field-Effect Transistors Made of an Individual Nitrogen-Doped Multiwalled Carbon Nanotube. *J. Am. Chem. Soc.* **2005**, *127*, 8614-8617.
58. Wang, X.; Li, X.; Zhang, L.; Yoon, Y.; Weber, P. K.; Wang, H.; Guo, J.; Dai, H., N-Doping of Graphene through Electrothermal Reactions with Ammonia. *Science* **2009**, *324*, 768-771.
59. Vitek, V., Intrinsic Stacking Faults in Body-Centred Cubic Crystals. *Philos. Mag.* **1968**, *18*, 773-786.
60. Christian, J.; Vitek, V., Dislocations and Stacking Faults. *Rep. Prog. Phys.* **1970**, *33*, 307-411.
61. Peierls, R., The Size of a Dislocation. *Proc. Phys. Soc.* **1940**, *52*, 34-37.
62. Nabarro, F., Dislocations in a Simple Cubic Lattice. *Proc. Phys. Soc.* **1947**, *59*, 256-272.
63. Joos, B.; Ren, Q.; Duesbery, M., Peierls-Nabarro Model of Dislocations in Silicon with Generalized Stacking-Fault Restoring Forces. *Phys. Rev. B* **1994**, *50*, 5890 1-9.
64. Juan, Y.-M.; Kaxiras, E., Generalized Stacking Fault Energy Surfaces and Dislocation Properties of Silicon: A First-Principles Theoretical Study. *Philos. Mag. A* **1996**, *74*, 1367-1384.
65. Joos, B.; Duesbery, M., The Peierls Stress of Dislocations: An Analytic Formula. *Phys. Rev. Lett.* **1997**, *78*, 266 1-4.
66. Hartford, J.; Von Sydow, B.; Wahnström, G.; Lundqvist, B., Peierls Barriers and Stresses for Edge Dislocations in Pd and Al Calculated from First Principles. *Phys. Rev. B* **1998**, *58*, 2487 1-10.
67. Lu, G.; Kioussis, N.; Bulatov, V. V.; Kaxiras, E., Generalized-Stacking-Fault Energy Surface and Dislocation Properties of Aluminum. *Phys. Rev. B* **2000**, *62*, 3099 1-10.
68. Hohenberg, P.; Kohn, W., Inhomogeneous Electron Gas. *Phys. Rev. B* **1964**, *136*, 864 1-8.
69. Kohn, W.; Sham, L. J., Self-Consistent Equations Including Exchange and Correlation Effects. *Phys. Rev. A* **1965**, *140*, 1133 1-6.
70. Kresse, G.; Hafner, J., Ab Initio Molecular Dynamics for Liquid Metals. *Phys. Rev. B* **1993**, *47*, 558 1-4.
71. Reich, S.; Maultzsch, J.; Thomsen, C.; Ordejon, P., Tight-Binding Description of Graphene. *Phys. Rev. B* **2002**, *66*, 035412 1-5.
72. Reich, S.; Thomsen, C.; Ordejon, P., Elastic Properties of Carbon Nanotubes under Hydrostatic Pressure. *Phys. Rev. B* **2002**, *65*, 153407 1-4.
73. Jun, S.; Li, X.; Meng, F.; Ciobanu, C. V., Elastic Properties of Edges in Bn and Sic Nanoribbons and of Boundaries in C-Bn Superlattices: A Density Functional Theory Study. *Phys. Rev. B* **2011**, *83*, 153407 1-4.
74. Ariza, M.; Serrano, R.; Mendez, J.; Ortiz, M., Stacking Faults and Partial Dislocations in Graphene. *Philos. Mag.* **2012**, *92*, 2004-2021.
75. Note: (cf. Figure 1) the smallest dimension corresponds to the 100% percent of impurity B or N concentration; the largest dimension corresponds to the 20% percent of impurity concentration.
76. Perdew, J. P.; Burke, K.; Ernzerhof, M., Generalized Gradient Approximation Made Simple. *Phys. Rev. Lett.* **1996**, *77*, 3865 1-4.
77. Blöchl, P. E.; Jepsen, O.; Andersen, O. K., Improved Tetrahedron Method for Brillouin-Zone Integrations. *Phys. Rev. B* **1994**, *49*, 16223 1-12.

78. Pulay, P., Convergence Acceleration of Iterative Sequences. The Case of Scf Iteration. *Chem. Phys. Lett.* **1980**, *73*, 393-398.
79. Brito, W.; Kagimura, R.; Miwa, R., B and N Doping in Graphene Ruled by Grain Boundary Defects. *Phys. Rev. B* **2012**, *85*, 035404 1-6.
80. Song, J.; Ouyang, B.; Medhekar, N. V., Energetics and Kinetics of Li Intercalation in Irradiated Graphene Scaffolds. *ACS Appl. Mater. Interfaces* **2013**, *5*, 12968-12974.
81. Singh, R.; Kroll, P., Magnetism in Graphene Due to Single-Atom Defects: Dependence on the Concentration and Packing Geometry of Defects. *J. Phys.: Condens. Matter* **2009**, *21*, 196002 1-7.
82. Xu, M.; Tabarraei, A.; Paci, J. T.; Oswald, J.; Belytschko, T., A Coupled Quantum/Continuum Mechanics Study of Graphene Fracture. *Int. J. Fract.* **2012**, *173*, 163-173.
83. Note: here for impurity-doped graphene, the cases with dopant concentration = 100% are presented for easy visualization.
84. Note: here the absolute bond length is measured without considering its angle with the shear direction.
85. Note: here for impurity-doped graphene, the cases with dopant concentration = 100% are presented for accurate calculations of pair distance.
86. Hirth, J. P.; Lothe, J., *Theory of Dislocations*; Krieger Publishing Company: Florida, U.S.A., 1982.
87. Note: the equilibrium separation between the partials can be further increased by the co-existence of N along the slip and B neighboring the slip doping, to as large as 1.3b, as shown in the Supporting Information.
88. Telling, R. H.; Heggie, M. I., Stacking Fault and Dislocation Glide on the Basal Plane of Graphite. *Philos. Mag. Lett.* **2003**, *83*, 411-421.
89. Huntington, H., Modification of the Peierls-Nabarro Model for Edge Dislocation Core. *Proc. Phys. Soc., Sect. B* **1955**, *68*, 1043 1-6.
90. Tuinstra, F.; Koenig, J. L., Raman Spectrum of Graphite. *J. Chem. Phys.* **1970**, *53*, 1126-1130.
91. Ertekin, E.; Chrzan, D.; Daw, M. S., Topological Description of the Stone-Wales Defect Formation Energy in Carbon Nanotubes and Graphene. *Phys. Rev. B* **2009**, *79*, 155421 1-17.
92. Yip, S., *Handbook of Materials Modeling*; Springer: Berlin, Germany, 2005.
93. Min, K.; Aluru, N., Mechanical Properties of Graphene under Shear Deformation. *Appl. Phys. Lett.* **2011**, *98*, 013113 1-3.

**FOR TABLE OF CONTENTS ONLY**

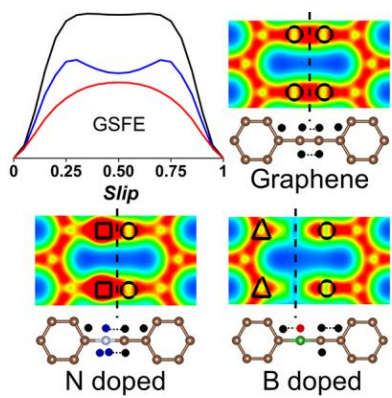

(For Table of Contents Only)